\documentclass[epjST]{svjour}
\usepackage[sort&compress,numbers,merge]{natbib}
\usepackage{amsmath}
\usepackage{amssymb}
\usepackage{amsfonts}
\usepackage{graphicx}
\usepackage{color}
\usepackage{subcaption}
\graphicspath{{images/}}
\usepackage{microtype}
\usepackage{hyperref} 
\newcommand{\unit}{1\!\!1}

\definecolor{DarkBlue}{RGB}{10,10,140}

\begin{document}

\title{The pseudogap regime in the unitary Fermi gas}
\author{S.~Jensen,$^1$ C. N.~Gilbreth,$^2$ and Y. Alhassid$^1$}
\institute{$^{1}$Center for Theoretical Physics, Sloane Physics Laboratory, Yale University, New Haven, CT 06520\\
$^{2}$Institute for Nuclear Theory,
Box 351550, University of Washington, Seattle, WA 98195}
\date{\today}

\abstract{We discuss the pseudogap regime in the cold atomic unitary Fermi gas, with a particular emphasis on the auxiliary-field quantum Monte Carlo (AFMC)  approach. We discuss possible signatures of the pseudogap, review experimental results, and survey analytic and quantum Monte Carlo techniques before focusing on AFMC calculations in the canonical ensemble. For the latter method, we discuss results for the heat capacity, energy-staggering pairing gap, spin susceptibility, and compare to experiment and other theoretical methods.}
\maketitle
\section{Introduction}
The unitary Fermi gas (UFG) describes a system of spin-1/2 particles interacting through a zero-range interaction tuned to the limit of infinite scattering length. It is a strongly interacting many-particle system with connections to high-$T_c$ superconductors~\cite{Randeria1992,Trivedi1995,Randeria2010},  quark matter~\cite{Nishida2005},  QCD plasmas~\cite{Adams2012} and neutron stars~\cite{Gandolfi2015}, and exhibits a superfluid phase transition with a high critical temperature $T_c$ in units of the Fermi temperature $T_F$.

The UFG sits midway in the BCS-BEC crossover, the continuous transition between degenerate Fermi and Bose gases obtained by varying the parameter $(k_{F}a)^{-1}$, where $a$ is the s-wave scattering length and $k_{F}$ is the Fermi momentum.  The Bardeen-Cooper-Schrieffer (BCS) regime, obtained for $(k_F a)^{-1} \sim - \infty$, is well-described by the BCS theory of Cooper pairs in the superfluid regime and by normal Fermi liquid theory for $T>T_{c}$. The Bose-Einstein condensate (BEC) regime, corresponding to  $(k_F a)^{-1} \sim +\infty$, consists of a weakly interacting Bose gas of tightly bound dimers with binding energy $E= -\hbar^2/(m a^2)$. In the unitary limit $(k_F a)^{-1} = 0$, the two-body bound state becomes a zero-energy resonance and the $s$-wave scattering cross section is maximized. 
The BCS-BEC crossover has been realized experimentally with ultra-cold atomic Fermi gases of $^{6} \textrm{Li}$ and $^{40} \textrm{K}$ near broad Feshbach resonances (for reviews see Refs.~\cite{Bloch2008,Giorgini2008}).

The unitary regime is highly non-perturbative and presents a major challenge to theorists.
The importance of the UFG is highlighted by the problem of the pseudogap regime in high-$T_{c}$ superconductors (e.g., the underdoped cuprates), in which a gapped structure persists above the superfluid critical temperature, but the exact mechanism responsible has eluded a precise theoretical description~\cite{Randeria2010,Dagotto1994,Timusk1999}. The difficulty in understanding the pseudogap regime in these high-$T_{c}$ superconductors provides a strong motivation for understanding pairing correlations above $T_c$ within the simpler context of the UFG, in which pseudogap effects can occur via pre-formed pairs~\cite{Randeria2010}. The pseudogap regime in the UFG itself has generated considerable debate about whether  it exists and what is its temperature range~\cite{Mueller2017}. Resolving this, from a theoretical perspective, requires precise theoretical calculations with controllable errors.  

In this brief review, we  discuss important theoretical and experimental results related to the pseudogap problem in the UFG, with a focus on our own recent quantum Monte Carlo simulations~\cite{Jensen2018}.  Pseudogap physics in the UFG has also been discussed elsewhere~\cite{Mueller2017,Chen2005,Chen2014}. 

This review is organized as follows. In Sec.~\ref{unitary} we introduce the UFG and the continuum and lattice models used to study this system. In Sec.~\ref{obs} we discuss possible signatures of the pseudogap regime in the UFG. In Sec.~\ref{experiment} we review landmark experimental results, and in Sec.~\ref{theory} we present an overview of theoretical approaches, emphasizing results related to the pseudogap regime.  In Sec.~\ref{Canonical} we discuss our recent canonical-ensemble quantum Monte Carlo simulations and their implications for the pseudogap. In Sec.~\ref{conclusion} we conclude and discuss future prospects.  

\section{Unitary Fermi gas}\label{unitary}

\subsection{Scattering and the unitary limit}

Neutral cold atomic gases interact at large distances via the Van der Waals interaction $V=-C/r^6$, effectively a short-range interaction. At low temperatures and densities, their scattering properties are well described by the low-momentum s-wave expansion for the scattering amplitude
\begin{equation}
f(k)=\frac{1}{-1/a -ik +r_{e}k^{2}/2 + ...} \;,
\end{equation}
where $a$ is the $s$-wave scattering length, $r_{e}$ is the effective range, and $...$ denotes higher-order corrections in the relative momentum $k$.  This expansion is justified for neutral cold atomic interactions in the limits $r_{0}\ll 1/k_{F}, \lambda_{T}$ for interaction range $r_{0}$ and thermal de Broglie wavelength $\lambda_{T}=\sqrt{2\pi\hbar^{2}/mk_{B}T}$. Here $k_{F}=(3\pi^{2}n)^{1/3}$ is the Fermi momentum for particle density $n$, and $k_B$ is the Boltzmann constant.  For a broad Feshbach resonance, the effective range parameter $r_{e}$ is of the same order as $r_{0}$~\cite{Castin2012}.

The unitary limit is obtained when the modulus of the scattering amplitude $|f(k)|$ is maximized, constrained by unitarity. From the optical theorem with s-wave scattering, $\textrm{Im} f(k) = k|f(k)|^{2}$, giving $f(k)=i/k$. The unitary limit is therefore characterized by the conditions $r_{e}\ll 1/k_{F}, \lambda_{T}$ and $k_{F}a\rightarrow \infty$.

\subsection{Continuum Model}

In the vicinity of a broad Feshbach resonance, the lowest two atomic hyperfine states of the atoms can be modeled by two effective spin-$1/2$ states that interact via a contact interaction.
The Hamiltonian of this system can be written as
\begin{equation}\label{Ham}
\hat{H}=-\sum_{\sigma} \int d^3 \bold{r} \hat{\psi}^{\dagger}_{\sigma}(\bold{r})\frac{\hbar^2 \bigtriangledown^{2}}{2m }\hat{\psi}_{\sigma}(\bold{r}) +\int d^3 \bold{r} \hat{\psi}^{\dagger}_{\uparrow}(\bold{r})  \hat{\psi}_{\downarrow}(\bold{r}) V(\bold{r}-\bold{r}') \hat{\psi}^{\dagger}_{\downarrow}(\bold{r}') \hat{\psi}_{\uparrow}(\bold{r}') \;,
\end{equation}
where $\hat{\psi}^{\dagger}_{\sigma}(\bold{r})$ [$\hat{\psi}_{\sigma}(\bold{r})$] creates (annihilates) a fermion at position $\bold{r}$ with spin $\sigma$, obeying anti-commutation relations $\{  \hat{\psi}_{\sigma}(\bold{r}),\hat{\psi}^{\dagger}_{\sigma'}(\bold{r}')\}=\delta(\bold{r}-\bold{r}')\delta_{\sigma,\sigma'}$. In coordinate space an appropriate potential is the regularized pseudopotential $V({\bf r})= g \delta(\bold{r}) \frac{\partial}{\partial r} r$, with $g=4\pi\hbar^2a/m$~\cite{Busch1998,Giorgini2008}. One can also use a regularized contact interaction $V(r) = V_0 \delta(\bold{r}-\bold{r'})$  with a bare coupling $V_{0}$ chosen to produce the desired two-body scattering length. $V_0$ is determined from the Lippmann-Schwinger equation, leading to~\cite{Werner2012}
\begin{equation}\label{renorm}
\frac{1}{V_{0}}=\frac{m}{4\pi \hbar^2 a}-\int^{\Lambda}\frac{d^{3}k}{(2\pi)^{3}2\epsilon_{k}}
\end{equation}
with hard ultraviolet momentum cutoff $\Lambda$.  Eq.~(\ref{Ham}) is the most frequently used model in theoretical studies of the pseudogap. Exact results are available for small numbers of particles~\cite{Werner2007}.

\subsection{Lattice Model}\label{lattice}

The Hamiltonian~\eqref{Ham} for a uniform gas can be approximated on a cubic lattice with $\textrm{M}=N_L^3$ lattice sites, lattice spacing $\delta x$, and periodic boundary conditions via
\begin{equation} \label{ham}
\hat{H}=\sum_{\bf{k},\sigma }\epsilon _{\bf{k}}\hat{a}^{\dagger }_{\bf{k},\sigma }\hat{a}_{\bf{k},\sigma }+g\sum_{\bf{x}}\hat{n}_{\bf{x},\uparrow}\hat{n}_{\bf{x},\downarrow} \;,
\end{equation}
where the single-particle dispersion is usually taken to be either quadratic $\epsilon _{\bf{k}} = {\hbar^2 \bf{k}}^{2} /2m$ or of the tight-binding form $\epsilon _{\bf{k}} = \frac{\hbar^2}{m\delta x^2}\sum_{i=1}^{3}\left[1-\textrm{cos}\left(k_{i}\delta x\right)\right]$ (the subscript $i$ labels the momentum component), and $g=V_0/(\delta x)^{3}$ is the lattice coupling.  The operators $\hat{a}^{\dagger }_{\bf{k},\sigma}$ and $\hat{a}_{\bf{k},\sigma}$ are, respectively, creation and annihilation operators in a single-particle state with wavevector ${\bf k}$ and spin $\sigma=\pm 1/2$, and $\hat{n}_{\bf{x},\sigma}=\hat{\psi}^{\dagger}_{\bf{x},\sigma}\hat{\psi}_{\bf{x},\sigma}$ are on-site number operators.  The lattice site creation and annihilation operators $\hat{\psi}^{\dagger}_{\bf{x},\sigma}, \hat{\psi}_{\bf{x},\sigma}$ obey the anti-commutation relations $\{ \hat{\psi}_{\bf{x},\sigma},\hat{\psi}^{\dagger}_{\bf{x}',\sigma'}\}= \delta_{\bf{x},\bf{x}'}\delta_{\sigma,\sigma'}$.

As discussed in Refs.~\cite{Werner2012,Jensen2018}, the single-particle model space should include all states within the complete first Brillouin zone, described by a cube in momentum space $|k_i| \le \Lambda \,\,\, (i=x,y,z)$ with $\Lambda=\pi/\delta x$. The bare coupling constant  $V_0$ is chosen to reproduce the two-particle scattering length as in Eq.~\eqref{renorm} where the integration region corresponds to the complete first Brillouin zone. The UFG is reproduced from the lattice model in the continuum limit $\nu\equiv N/N_L^3\rightarrow 0$ and thermodynamic limit $N\rightarrow \infty$.

\subsection{Trapped gases}\label{trapped}

Trapped gases can be described by Eq.~\eqref{Ham} with the addition of a trapping potential $V_{\rm trap}(r)$. We have considered an isotropically trapped system with $V_{\rm trap}(r) = m \omega^2 r^2 / 2$~\cite{Gilbreth2013}. A natural single-particle basis is given by the eigenstates $|n,l,m_l\rangle$ of $H_0 = -\frac{\hbar^2}{2 m} \nabla^2 + V_{\rm trap}$, where $n$ is the radial quantum number, $l$ is the orbital angular momentum, and $m_l$ the magnetic quantum number, with energy $\epsilon_{n l} = (2n+l)\hbar \omega$. One can truncate the single-particle basis to a total of $N_{\rm max}$ quanta of energy, i.e., $2 n + l \leq N_{\rm max}$. An effective interaction was developed for trapped systems with fast convergence in a regularization parameter~\cite{Alhassid2008,Gilbreth2012}, but it does not have a good Monte Carlo sign. We consider only a contact interaction $V(r) = V_0 \delta(\bold{r}-\bold{r'})$, with $V_0$ tuned to reproduce the two-particle ground-state energy as calculated in Ref.~\cite{Busch1998}. We note this regularization is only approximate, the methods of Ref.~\cite{Stetcu2007,Alhassid2008,Gilbreth2012} being more rigorous. This interaction was also used in Ref.~\cite{Mukherjee2013}.

\section{Signatures of a pseudogap}\label{obs}

Different observables have been proposed as possible signatures of a pseudogap regime, and there is no one  definition which is universally accepted. Most often, the pseudogap regime refers to a depression in the single-particle density of states $\rho(\omega)$ at the chemical potential $\mu$ for temperatures above $T_c$. The single-particle density of states is given by $\rho(\omega)=\int \frac{d^{3} k}{(2\pi)^3}A(\bold{k},\omega)$, where $A(\bold{k},\omega)$ is the spectral weight.

\subsection{Spectral weight}

More generally, one can study features of the spectral weight $A(\bold{k},\omega)$ itself. In BCS theory, it has the form~\cite{Fetter1971,Haussmann2009}
\begin{equation}
A(\bold{k},\omega) = u_{\bold k}^2 \delta(\omega - E_{\bold k}^{(+)}) + v_{\bold k}^2 \delta(\omega - E_{\bold k}^{(-)}) \;,
\end{equation}
where $E_{\bold k}^{(\pm)} = \mu \pm E_{\bold{k}}$ 
and $E_{\bold{k}}=\sqrt{(\epsilon_{\bold k} - \mu)^2 + \Delta^2}$ is the quasiparticle energy ($\Delta$ is the single-particle excitation gap and $\epsilon_{\bold{k}}=\frac{\hbar^2 \bold{k}^2}{2m}$).  The amplitudes $u_{\bold k},v_{\bold k}$  define the quasiparticle creation operators $\hat{\alpha}^\dagger_{\bold{k} \uparrow}=u_{\bold{k}}\hat{a}^\dagger_{\bold{k}\uparrow}-v_{\bold{k}}\hat{a}_{-\bold{k} \downarrow}$ and $\hat{\alpha}^\dagger_{-{\bold{k} \downarrow}}=u_{\bold{k}}\hat{a}^\dagger_{-\bold{k}\downarrow}+v_{\bold{k}}\hat{a}_{\bold{k} \uparrow}$ with $u_{\bold{k}}^2+v_{\bold{k}}^2=1$.
In a strongly correlated system, which cannot be well described by mean-field theory, $A(\bold{k},\omega)$ will have a more complex form with broadened peaks. However, these broadened peaks can sometimes be fit for $T > T_c$ to a modified quasiparticle dispersion~\cite{Magierski2009,Haussmann2009}
\begin{equation}\label{eq:modified_dispersion}
E_{\bold{k}}^{(\pm)}=\mu \pm \sqrt{\left(\frac{m}{m^{*}}\epsilon_{k}+U-\mu\right)^2+\Delta^2} \;.
\end{equation}
Here $m^{*}$ is the effective mass and $U$ is a Hartree shift parameter.  A nonzero $\Delta$ for $T>T_{c}$ indicates the existence of gapped quasiparticle excitations due to pre-formed pairs.  However, a more general signature of a pseudogap is a suppression of $A({\bf k},\omega)$ near the chemical potential $\omega=\mu$.

One can define a temperature scale $T^*$ as the temperature at which signatures of pairing first appear as the temperature is lowered. Note that in general no phase transition occurs at $T^*$.
In BCS theory, the onset of pairing and condensation occur simultaneously and $T^*=T_c$, where the gap parameter $\Delta$ is the order parameter describing the phase transition. In the BEC regime, however, the pairing temperature $T^*$ is the temperature scale associated with the formation of dimers and is distinct from the much lower condensation temperature $T_{c}$.  For the UFG, 
the exact value of $T^*$ ($T^* \geq T_c$) is still debated in the literature.

\subsection{Thermodynamics}

Several thermodynamic quantities have been studied for signatures of pseudogap effects.\\

\noindent (i) {\em Heat capacity}. The heat capacity $C_V= (\partial E/\partial T)_V$ can be affected by pseudogap physics.  Underdoped cuprate high-$T_c$ superconductors, which display pseudogap effects, are seen to exhibit a suppression of $\gamma=C_V / T$ for $T > T_c$~\cite{Timusk1999}, indicating deviation from Fermi liquid theory.  On the other hand, it has been argued that in the UFG the specific heat should be enhanced above $T_c$ (either relative to the BCS and BEC regimes, or by showing an upward trend as $T$ approaches $T_c$ from above) due to the existence of $T>T_{c}$ precursor pairing correlations~\cite{Strinati2011,Ohashi2016}. The specific heat was measured across the superfluid phase transition in the UFG~\cite{Ku2012} (see Sec.~\ref{experiment}).\\

\noindent (ii) {\em Spin susceptibility}. The uniform static spin susceptibility $\chi_s$
\begin{equation} \label{susceptibility}
\chi_{s}=\frac{1}{k_B T V}\langle(\hat{N}_{\uparrow}-\hat{N}_{\downarrow})^{2}\rangle \;,
\end{equation}
also provides a signature of pseudogap effects as discussed in Refs.~\cite{Randeria1992,Trivedi1995}.  In this case, pairing correlations tend to suppress $\chi_s$ as pair breaking excitations become energetically unfavorable. The spin susceptibility is therefore expected to drop below $T^*$, similar to the exponential suppression of the spin susceptibility for a BCS superfluid at low temperatures.\\

\noindent (iii) {\em Pairing gap}. A model-independent pairing gap can be defined from the staggering of the energy with particle number
\begin{equation}\label{gap}
\Delta_E \! = \! [2E(N/2+1,N/2)-E(N/2+1,N/2+1)-E(N/2,N/2)]/2 \,,
\end{equation} 
where $E(N_1,N_2)$ denotes the energy of a system with $N_1$ spin-up and $N_2$ spin-down particles, and $N=N_1+N_2$ is the total number of particles~\cite{Carlson2003,Giorgini2008}. This definition is used for atomic nuclei~\cite{Bohr1969} at zero temperature. We apply Eq.~\eqref{gap} more generally at finite temperature.  Note that $E(N_1,N_2)$ in Eq.~\eqref{gap} is defined in the canonical ensemble of fixed particle numbers $N_1$ and $N_2$. This has been implemented in AFMC using particle-number projection~\cite{Gilbreth2013,Jensen2018} and reprojection~\cite{Alhassid1999}, as discussed in Sec.~\ref{Canonical}.\\

\noindent (iv) \emph{Equation of state}. The equation of state has been studied in the normal phase and compared with predictions of Fermi liquid theory. In particular, the pressure of a Fermi liquid is expected to vary linearly with $(k_B T/\mu)^2$, when properly normalized.  This has been investigated experimentally as discussed in Sec.~\ref{sec:experiment} below.\\ 

\noindent (v) {\em Contact}. The temperature dependence of Tan's contact~\cite{Tan2008,Tan2008-2,Tan2008-3} has also been suggested to exhibit signatures of the pseudogap~\cite{Palestini2010,Drut2011}. The contact is the coefficient of the $k^{-4}$ tail of the momentum distribution $n_{\sigma}(\bold{k})$ for large $k$, and as such is related to short-distance properties of the gas. 
See also Refs.~\cite{Pieri2009,Sagi2012,Schneider2010}.

\section{Experimental results}\label{experiment}

\label{sec:experiment}

Effects of pairing were observed in early spectroscopic measurements of trapped Fermi gases~\cite{Chin2004,Greiner2005}. These experiments did not directly address pseudogap physics above $T_c$, but provided the first experimental evidence of pairing in the unitary regime. 

The 2010 experiment of Ref.~\cite{Gaebler2010} appears to have been the first attempt to directly address pseudogap physics by measuring the spectral function of a trapped gas.
A BCS-like dispersion, with backbending near $k_F$, was observed in photoemission spectra at temperatures greater than $T_c$. These results were subsequently analyzed in several papers with contradictory conclusions being drawn, one group pointing to a Fermi-liquid behavior~\cite{Navon2011}, and another explaining them in terms of a pseudogap~\cite{Strinati2011}.

\begin{figure}[htb]
\begin{minipage}[t]{0.45\textwidth}
\centering
\includegraphics[width=0.75\textwidth]{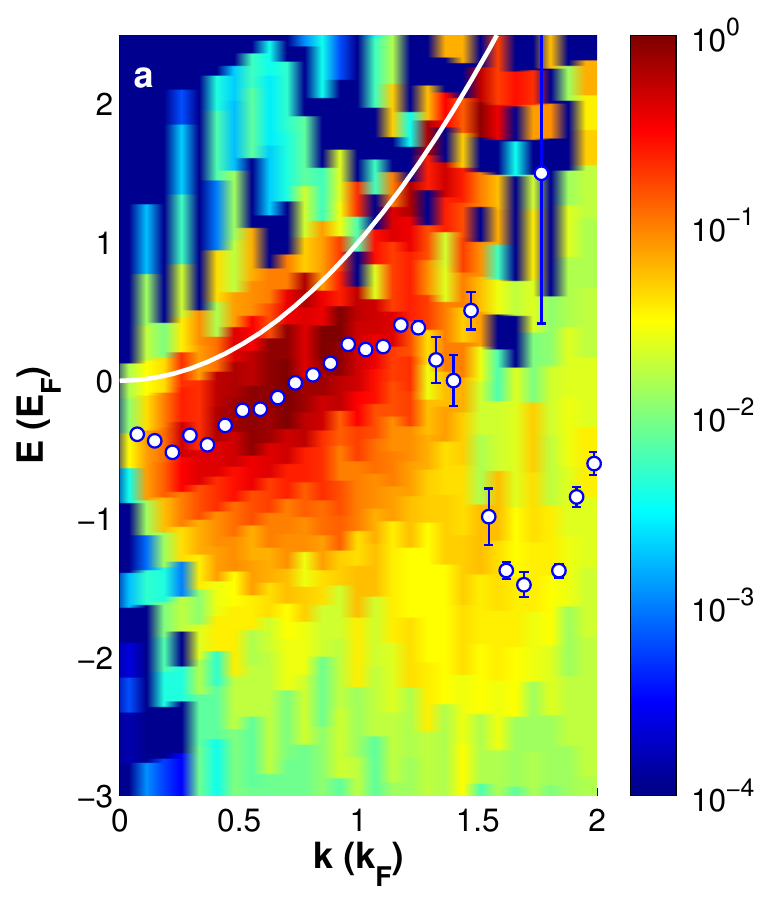}
\caption{Photoemission spectroscopy (PES) data~\cite{Jin2015} for the homogenous Fermi gas near unitarity [$(k_F a)^{-1} = 0.1$] at $T > T_c$. Color shows the normalized PES signal $I(k,E)$ (see text) as a function of the single-particle momentum $k$ and energy $E$. The white line indicates the free particle dispersion $E \propto k^2$, and the circles are the peak positions. Adapted from the supplementary material of Ref.~\cite{Jin2015}.}
\label{fig:Jin2015_spectral}
\end{minipage}
\hspace{0.25cm}
\begin{minipage}[t]{0.45\textwidth}
\centering
\includegraphics[scale=0.7]{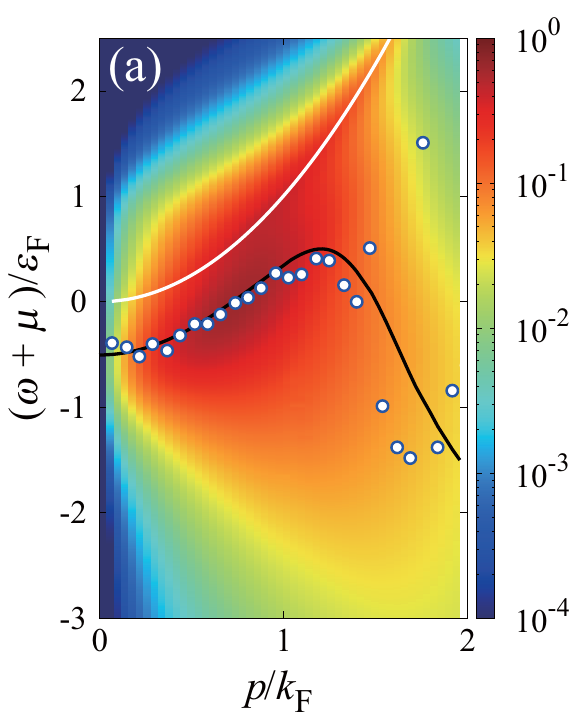}
  \caption{Spectral intensity near unitarity [$(k_F a)^{-1} = 0.1$] computed using a $T$-matrix approximation and compared with the experimental data in Fig.~\ref{fig:Jin2015_spectral}. Color shows the computed intensity $I(k,E)$ (see text) including effects of the trap. The calculated peak position (black solid line) is compared to the experimental peak positions (white circles). Adapted from Ref.~\cite{Ohashi2017}.}
\label{fig:Ohashi2017_spectral}
\end{minipage}
\end{figure}

A more recent measurement accounted for trapping effects by focusing attention to the center of the cloud~\cite{Jin2015}. This experiment measured $I(k,E)\propto k^2A(k,E)f(E)$, where $f(E)$ is the Fermi function and $A(k,E)$ is the single-particle spectral function. The result is shown in Fig.~\ref{fig:Jin2015_spectral}, normalized so that $\int dk \, dE\, I(k,E) = 1$, near unitarity [$(k_F a)^{-1} = 0.1$] and slightly above the critical temperature. The experimental data, including effects of finite resolution, is well-modeled by a fermionic quasiparticle spectral function plus an incoherent background consisting of a thermal distribution of bound pairs, with the weight of the quasiparticle signal decaying to zero beyond $k_F a \approx 0.28$. 
This data has been analyzed favorably in terms of a $T$-matrix theory which displays a pseudogap~\cite{Ohashi2017} (see Sec.~\ref{strong-coupling} and Fig.~\ref{fig:Ohashi2017_spectral}).

\begin{figure}[htb]
\begin{center}
\includegraphics[width=0.6\textwidth]{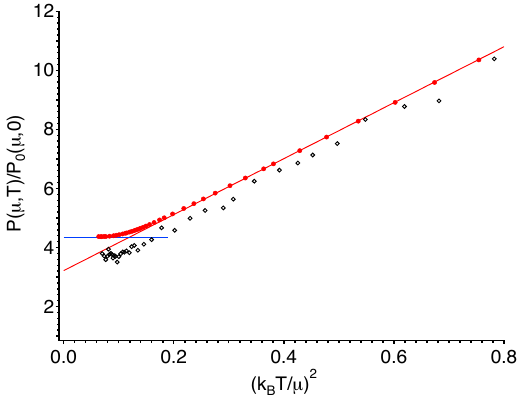}
\end{center}
\caption{Pressure of the homogeneous Fermi gas at unitarity as a function of $(k_B T/\mu)^2$. The vertical axis shows the local pressure $P(\mu,T)$ of the trapped gas, normalized by the pressure $P_0(\mu,0)$ of a non-interacting Fermi gas at zero temperature. The red solid circles are the experiment of Ref.~\cite{Ku2012}, the red line is a linear fit, and the solid blue line is the zero-temperature limit. Open circles are the earlier experimental results from Ref.~\cite{Navon2009}. The offset  between the two experimental results is not fully understood  but might be  in part due to uncertainties in the Feshbach resonance position~\cite{NavonPrivateComm}. The linear behavior above $T \sim T_c$ is as expected for a Fermi liquid. Adapted from the supplemental material of Ref.~\cite{Ku2012}.}
\label{fig:expt_thermo}
\end{figure}

A measurement of the equation of state in 2010 found a Fermi-liquid type behavior of the pressure as a function of temperature~\cite{Navon2010}. The magnetic susceptibility in the normal phase was subsequently determined~\cite{Navon2011} and shown to be consistent with a Fermi-liquid behavior. However, shortly afterward a measurement of the specific heat $C_V$ across the superfluid phase transition~\cite{Ku2012}, indicated non-Fermi-liquid behavior, in that $C_V$ was not found to be linear in $T$ in the normal phase. The same experiment also found a Fermi-liquid-like behavior in the pressure above $T_c$; see Fig.~\ref{fig:expt_thermo}.

In Ref.~\cite{Sommer2011} the spin susceptibility was determined as a function of temperature, and no downturn in the spin susceptibility was observed at lower temperatures. This result though, is not in good quantitative agreement with current theoretical results~\cite{Enss2012,Strinati2012-2}.

The contact was measured in Ref.~\cite{Sagi2012}.  It did not show a peak in the normal phase as predicted by strong-coupling theories~\cite{Palestini2010,Hu2011}, but exhibited a significant drop for temperatures below $T_c$.

\section{Theoretical approaches}\label{theory}

\subsection{Strong-coupling theories} \label{strong-coupling}

Early analytical approaches to the BCS-BEC crossover include those of Eagles~\cite{Eagles1969} and Leggett~\cite{Leggett1980}, and Nozieres and Schmidt-Rink (NSR)~\cite{Nozieres1985}.
NSR included pairing fluctuations to determine the dependence of $T_c$ on the interaction strength. Ref.~\cite{Melo1993} obtained similar NSR equations for $T_c$ using Gaussian fluctuation theory.

Subsequently, various theories have emerged treating pairing correlations in the strong-coupling regime beyond the mean field by summing infinite series of certain diagrams in perturbation theory.  A general difficulty with these theories is that they include uncontrolled approximations whose errors cannot be estimated a priori. However, the two-species Fermi gas with contact interaction provides a useful testing ground for such theories, since it is experimentally accessible and permits quantum Monte Carlo simulations that are free of the sign problem.

A number of other theories addressing the pseudogap which we do not discuss here can be found in Refs.~\cite{DeWolfe2017,Stoof2011,Dupuis2016,Doggen2015}.

\paragraph{Non-self-consistent $T$-matrix approaches.} One of the earlier strong-coupling theories employed a non-self-consistent $T$-matrix approximation to describe the pseudogap in cold atomic Fermi gases~\cite{Strinati2002}. Similar methods were applied in Refs.~\cite{Strinati2011,Ohashi2009,Strinati2012,Reichl2015,Ohashi2017}. In particular, the spectral function has been calculated in Refs.~\cite{Ohashi2009,Strinati2011,Reichl2015} at unitarity and compared to experiment in Refs.~\cite{Strinati2011,Ohashi2017}. It agrees well with the recent experiment of Ref.~\cite{Jin2015} (see Fig.~\ref{fig:Ohashi2017_spectral} below).
The $T$-matrix approximation of Ref.~\cite{Strinati2002} predicts $T_c = 0.24~T_F$ at unitarity, while an extended version used in Refs.~\cite{Ohashi2012,Tajima2014,Tajima2014} predicts $T_c = 0.21~T_F$~\cite{Tajima2014}. A recent development of this method can be found in Ref.~\cite{Strinati2018}.

Another approach, which describes a finite-temperature extension of the work of Leggett, is discussed in Ref.~\cite{Levin2010} and referred to as the extended BCS-Leggett theory.  In comparison with that of Ref.~\cite{Strinati2002}, it is less accurate in the BEC regime, but more accurate in the BCS regime. 
It yields $T_c = 0.26~T_F$ at unitarity.  Refs.~\cite{Chien2010a,Levin2010} provide a detailed comparison between the extended BCS-Leggett theory and the $T$-matrix approximations (referred to as NSR theories in Refs.~\cite{Chien2010a,Levin2010}) discussed above.

\begin{figure}[htb]
\begin{center}
  \includegraphics[scale=0.75]{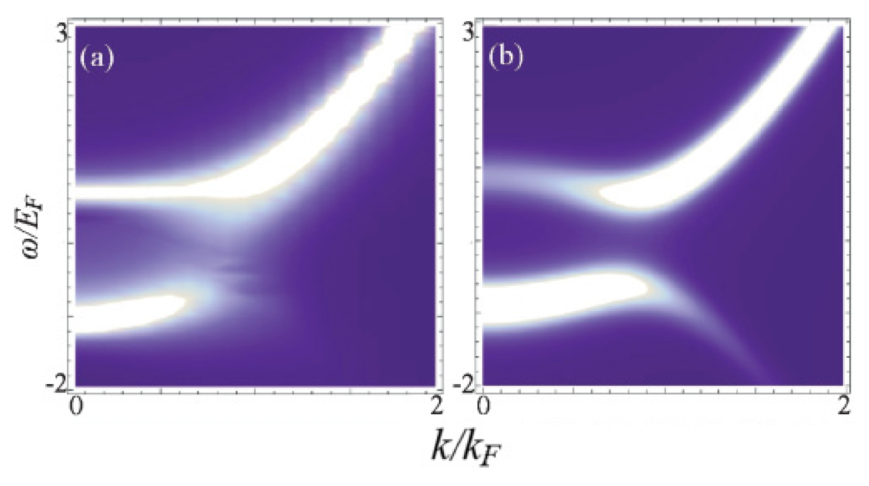}
  \caption{Comparison of the spectral function $A(k,E)$ of the UFG for two strong-coupling theories at $T=0.24\,T_F$ (which is at or slightly below $T_c$ for these theories). The left panel shows a $T$-matrix approximation similar to Ref.~\cite{Strinati2002}, and the right panel is obtained in the extended BCS-Leggett theory. Adapted from Ref.~\cite{Chien2010a}.}
\label{fig:Levin_spectral}
\end{center}
\end{figure}

Fig.~\ref{fig:Levin_spectral} shows a comparison from Ref.~\cite{Chien2010a} of the spectral function computed in the $T$-matrix approximation (left panel) and the extended BCS-Leggett theory (right panel), both at $T = 0.24 \,T_F$, which is at or slightly below $T_c$ for these theories.  Both theories show clear pseudogap effects, with the BCS-Leggett theory showing a more pronounced effect that includes backbending of the lower branch of the spectral function in addition to the two-peak structure. 

\begin{figure}
  \centering
  \begin{minipage}[t]{0.45\textwidth}
    \centering
    \includegraphics[width=.95\linewidth]{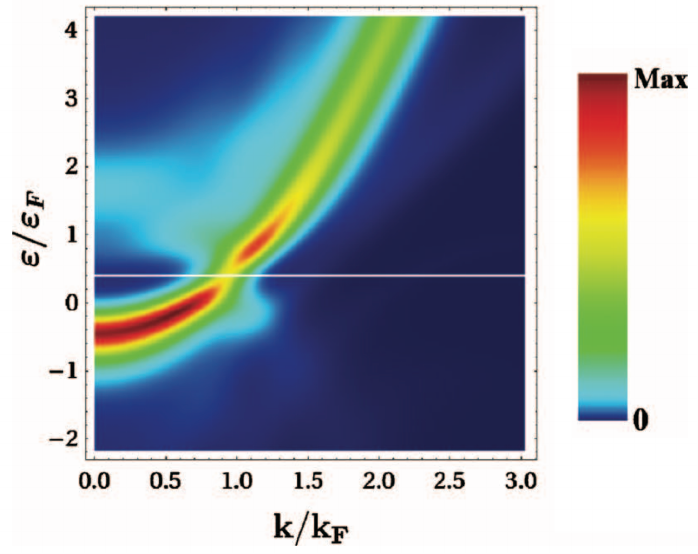}
    \caption{Spectral function $A (\mathbf{k}, \omega)$ at unitarity for $T=T_c$ 
      computed using a self-consistent Luttinger-Ward theory. At $T_c$, this theory
      shows only a weak two-branch structure. Adapted from Ref.~\cite{Haussmann2009}.}
    \label{fig:Haussmann_spectral}
  \end{minipage}
  \hspace{0.25cm}
  \begin{minipage}[t]{0.45\textwidth}
    \includegraphics[width=1.0\linewidth]{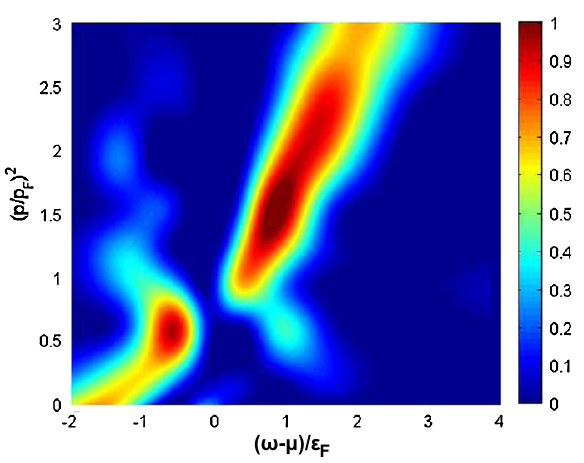}
    \caption{Spectral function $A(\bold{k},\omega)$ computed using AFMC at $T=0.15\,T_{F}$ (the critical temperature reported by the authors of this work is $T_{c}\lesssim0.15(1) \, T_{F}$~\cite{Bulgac2008}). The horizontal axis is $(\omega-\mu)/\epsilon_{F}$  and the vertical axis is $(p/p_{F})^{2}$, where $\mu$ is the chemical potential and $p_F$ is the Fermi momentum.  
       Adapted from Ref.~\cite{Magierski2009}.}
    \label{fig:Magierski}
  \end{minipage}
\end{figure}

\paragraph{A self-consistent Luttinger-Ward theory.} A self-consistent theory has been developed for the BCS-BEC crossover based on the Luttinger-Ward functional~\cite{Haussmann1993,Haussmann1994,Haussmann2007,Haussmann2009,Enss2012}.
In this approach $T_c = 0.16 (1)\, T_{F}$ at unitarity. When compared with other strong coupling theories, this value is closer to the experimental and quantum Monte Carlo values. 
Fig.~\ref{fig:Haussmann_spectral} shows the unitary gas spectral function calculated in this self-consistent approach at $T_c$.  It exhibits only weak evidence of a pseudogap effect. 

\subsection{Quantum Monte Carlo methods}

A number of quantum Monte Carlo approaches have been used to study the UFG and its properties above $T_c$.  There have also been many quantum Monte Carlo calculations of the UFG ground state ($T=0$) studying among others the Bertsch parameter~\cite{Carlson2005,Carlson2011}, Tan's contact~\cite{Gandolfi2011}, the superfluid pairing gap~\cite{Carlson2003,Carlson2008}, finite-size effects~\cite{Forbes2011}, and effective effective-range dependence~\cite{Li2011,Forbes2012,Schonenberg2017}.  Here we focus on finite-temperature approaches that are relevant to pseudogap phenomena.

\paragraph{Diagrammatic Monte Carlo methods}  The diagrammatic quantum Monte Carlo approach samples  stochastically the contributions of all Feynman diagrams to correlation functions in the interaction picture. This has been implemented on a lattice and used to calculate thermodynamic properties of the normal and superfluid phases, extrapolating to the continuum and thermodynamic limits. In particular the critical temperature~\cite{Burovski2006,Goulko2010}, and more recently the temperature dependence of the contact~\cite{Goulko2016} have been computed and compared with experiment. The lattice diagrammatic approach is  described in detail in Refs.~\cite{Rubstov2005,Burovski2006-2}. 

The ``bold diagrammatic'' approach~\cite{Prokofev2007,Prokofev2008,Van2012,Van2013-2} works with fully dressed Green's functions directly in the continuum and thermodynamic limits, and has seen very good agreement with experiment for thermodynamic properties of the UFG. To date this method has only been applied above $T_c$.

\paragraph{Auxiliary-field Monte Carlo method}

The auxiliary-field quantum Monte Carlo (AFMC) method~\cite{Blankenbecler1981,White1989} is based on the Hubbard-Stratonovich transformation~\cite{Stratonovich1957,Hubbard1959}, in which the thermal propagator $e^{-\beta \hat H}$ is written as a superposition of propagators of non-interacting particles in external   auxiliary fields. The sum over the auxiliary-field configurations is then sampled stochastically. Finite-temperature lattice AFMC in the grand-canonical ensemble has been used to calculate both thermodynamic and dynamic properties of the UFG~\cite{Bulgac2008,Magierski2009,Magierski2011,Drut2013-2}.
Unlike the diagrammatic methods, the computational cost of AFMC has so far prevented the extraction of observables in the full thermodynamic and continuum limits.

The spectral function has been computed for small lattices~\cite{Magierski2009,Magierski2011,Drut2013-2}.  Fig.~\ref{fig:Magierski} shows the result of Ref.~\cite{Magierski2009} for $N=50-55$ particles on a lattice of size  $8^3$ at $T=0.15 \, T_{F}$, which is essentially at the authors' estimated critical temperature $T_{c}\lesssim 0.15(1) \, T_{F}$. The result shows a clear gapped structure. A modified BCS dispersion~\eqref{eq:modified_dispersion} was fitted to extract an effective gap parameter, giving $\Delta \approx 0.22 \, \varepsilon_{F}$ at this temperature.

The static spin susceptibility was also computed~\cite{Drut2013-2} and seen to have substantial suppression for $T \ge T_c$ indicating a pseudogap regime where the pairing temperature scale is estimated to be $T^*=0.20-0.25T_{F}$.

We note that the calculations of Refs.~\cite{Bulgac2006,Bulgac2008,Magierski2009,Magierski2011,Drut2013-2} used a spherical cutoff in the single-particle space $|\bold{k}| \le \Lambda$ rather than the complete first Brillouin zone [see Eq.~(\ref{ham})].  As discussed in Refs.~\cite{Werner2012,Jensen2018}, this approximation affects the two-particle scattering and does not reproduce the scattering properties of the unitary gas even in the continuum limit. See also Sec.~\ref{cutoff}.

\paragraph{Dynamical cluster Monte Carlo approach}
The dynamical cluster quantum Monte Carlo approach~\cite{Hettler1998,Hettler2000,Maier2005} treats correlations on the lattice exactly for clusters of lattice sites up to a certain size, with longer-range correlations treated in a mean-field description.
 This approach has been applied to study the normal state spectral function with lattice density $\nu=0.3$ for several scattering lengths~\cite{Jarrell2010}. 
No clear pseudogap regime above the critical temperature was seen for the UFG.

\section{Canonical ensemble auxiliary-field quantum Monte Carlo methods}\label{Canonical}

We have developed AFMC methods in the canonical ensemble of fixed particle number for trapped Fermi gases of cold atoms~\cite{Gilbreth2013} and for uniform Fermi gases in a lattice approach~\cite{Jensen2018}. These approaches are inspired by AFMC methods developed for nuclei~\cite{Lang1993,Alhassid1994} in which the number of protons and neutrons are fixed; for reviews see Refs.~\cite{Koonin1997, Alhassid2001_intj, Alhassid2017}. An advantage of the canonical ensemble formulation is that it allows the computation of a pairing gap from the staggering of the energy in particle number, Eq.~(\ref{gap}). Effects of the canonical ensemble have also been discussed in Ref.~\cite{Assaad2017}. We use a particle-number projection method, whose direct implementation is computationally intensive. However, we have extensively optimized this projection technique so it does not increase significantly the overall computation time for most observables.

For our lattice calculations we use the model of Sec.~\ref{lattice} with fixed numbers of $N_{\uparrow}$ and $N_{\downarrow}$ fermions on discrete lattices with lattice spacing of $\delta x$ and $N_L$ points in each dimension (except when calculating the spin susceptibility $\chi_{s}$ where only the total number of particles $N=N_{\uparrow}+N_{\downarrow}$ is fixed). 
We use the single-particle dispersion relation $\epsilon_{\bold k} = \hbar^2 k^2/2m$ and include all single-particle momentum states $\hbar\bold{k}$ within the first full Brillouin zone.  For trapped gases we use the model of Sec.~\ref{trapped} with fixed numbers of $N_{\uparrow}$ and $N_{\downarrow}$ fermions.

\subsection{AFMC method}\label{AFMC_method}
The AFMC method is based on the Hubbard-Stratonovich transformation~\cite{Stratonovich1957,Hubbard1959} which represents the thermal propagator $e^{-\beta\hat{H}}$  as a functional integral over auxiliary fields of one-body propagators. We sketch below the derivation.

 A general two-body Hamiltonian can be written in the form $\hat{H}=\hat{H}_0+\hat{V}$ where $\hat H_0$ is a one-body operator and $\hat{V} = \sum_\alpha \lambda_\alpha \hat O_\alpha^2$ is the sum of squares of one-body operators $\hat O_\alpha$. In the lattice case, $\hat H_0 = \hat K - g \hat N/2$, where $\hat K$ is the kinetic term, $\lambda_\alpha = g/2$, and $\hat O_\alpha$ are the on-site number operators $\hat n(\mathbf{x})=\hat{n}_{\mathbf{x},\uparrow}+\hat{n}_{\mathbf{x},\downarrow}$. In the trapped case, the Hamiltonian can be decomposed in an angular-momentum-conserving formalism~\cite{GilbrethThesis,Koonin1997}.

The propagator $e^{-\beta\hat{H}}$ is factorized using a Trotter decomposition $e^{-\beta \hat{H}} = (e^{-\Delta \beta \hat H})^{N_\tau}$, where we have divided the imaginary time $\beta$ into $N_{\tau}$ time slices ($\Delta \beta = \beta/N_{\tau}$). For each time slice, we use
$e^{-\Delta \beta \hat H} = e^{-\Delta \beta \hat K/2}e^{-\Delta \beta \hat{V}}e^{-\Delta \beta \hat K/2} + O((\Delta \beta)^3)$ for our lattice calculations, and $e^{-\Delta \beta \hat H} = e^{-\Delta \beta \hat H_0} e^{-\Delta \beta \hat{V}} + O((\Delta \beta)^2)$ for our trapped calculations. The overall error is then $O((\Delta \beta)^{2})$ or $O(\Delta \beta)$, respectively.
Introducing an auxiliary field $\sigma_{\alpha}(\tau_{n})$ for each $\hat O_\alpha$ and at each time slice $n$, we express $\textrm{exp}(-\Delta \beta\lambda_\alpha \hat O_\alpha^2)$ as a Gaussian integral over $\sigma_{\alpha}(\tau_{n})$ to obtain the Hubbard-Stratonovich transformation
\begin{equation} \label{pathint}
e^{-\beta \hat{H}} \approx \int D[\sigma ]G_{\sigma }\hat{U}_{\sigma } \;,
\end{equation}
where $D\left [\sigma \right ]$ is the integration measure and $G_{\sigma}$ is a Gaussian weight.  $\hat U_\sigma$ is the many-particle propagator
$  \hat{U}_{\sigma}= \hat U_{N_\tau} \cdots \hat U_1$, 
a time-ordered product with each $\hat U_n$ being a product of exponentials of one-body operators.

The expectation value of an observable $\hat{O}$ can be written as a functional integral 
\begin{equation}\label{path_integral}
\langle \hat{O}\rangle = \frac{\int D\left [ \sigma \right ] \langle \hat{O}\rangle_{\sigma} W_{\sigma} \Phi_{\sigma}}{\int D\left [ \sigma \right ] W_{\sigma} \Phi_{\sigma}} \,,
\end{equation}  
where $\langle \hat{O}\rangle_{\sigma}=\textrm{Tr}(\hat{O}\hat{U}_{\sigma})/\textrm{Tr}(\hat{U}_{\sigma})$ is the expectation value of $\hat O$ for a given auxiliary-field configuration $\sigma$, $W_{\sigma}=G_{\sigma}|\textrm{Tr}(\hat{U}_{\sigma})|$ is a positive-definite weight, and $\Phi_{\sigma}=\textrm{Tr}(\hat{U}_{\sigma})/|\textrm{Tr}(\hat{U}_{\sigma})|$ is the Monte Carlo sign.  The high-dimensional integral in Eq.~(\ref{path_integral}) is calculated using Monte Carlo sampling, for which we use the Metropolis-Hastings algorithm~\cite{Metropolis1953,Hastings1970}, updating one time slice at a time. The traces in the equations  above are calculated in the canonical ensemble as discussed in the next section.  

The advantage the AFMC framework is that $\hat U_\sigma$ in Eq.~(\ref{pathint}) is a one-body propagator, so the expectation values of observables with respect to $\hat U_\sigma$ can be computed using matrix algebra in the single-particle space. In this single-particle space $\hat U_\sigma$ is represented by a chain of matrix products. For a large number of time slices, this product must be numerically stabilized~\cite{Koonin1997}. This stabilization is achieved by a QDR decomposition: the matrix ${\bf U}_\sigma$ representing $\hat U_\sigma$ is stored in a decomposed form ${\bf U}_\sigma = Q D R$ where $Q$ is unitary, $D$ is diagonal with positive entries, and $R$ is unit upper triangular.

\subsection{Particle-number projection}\label{number}
For each auxiliary field configuration we project onto fixed particle number using a discrete Fourier transform~\cite{Ormand1994}, with projection operator
\begin{equation}\label{particle-projection}
\hat{P}_{N_\sigma}=\frac{e^{-\beta \mu N_\sigma}}{M}\sum_{m=1}^{M}e^{-i\varphi _{m}N_\sigma}e^{(\beta \mu +i\varphi _{m})\hat{N}_\sigma}
\end{equation}
for $\sigma = \uparrow$ or $\downarrow$. Here $\varphi _{m}=\frac{2\pi m}{M}$, $M=N_L^3$ is the number of lattice points (in the lattice case) or the number of single-particle harmonic oscillator states (in the trapped case), and the chemical potential $\mu$ is introduced for numerical stability.  We then evaluate the traces above in the canonical ensemble using $\textrm{Tr}_{N_{\uparrow},N_{\downarrow}}(\hat{X})=\textrm{Tr}_{\textrm{GC}}(\hat{P}_{N_\uparrow}\hat{P}_{N_\downarrow}\hat{X})$ which then allows us to write the canonical ensemble trace as sum of grand-canonical traces (with complex chemical potentials).  The grand-canonical trace is calculated in the single-particle space for the one-body propagator $\hat{U}_{\sigma}$ using the relation 
\begin{equation}\label{determinant}
\textrm{Tr}_{\textrm{GC}}[e^{(\beta \mu +i\varphi_{m})\hat{N}}\hat{U}_\sigma]=\textrm{det}[\unit +e^{(\beta \mu +i\varphi _{m})}\mathbf{U}_\sigma] \;.
\end{equation}

The canonical projection requires $O(M^4)$ operations to compute, particularly when combined with numerical stabilization. A canonical ensemble algorithm allowing $O(M^3)$ scaling was introduced in Refs.~\cite{Gilbreth2013,Gilbreth2015}. For our lattice calculations we have further reduced the computational time using several methods which effectively reduce the dimension of the single-particle model space for a given field configuration. In particular, the $D$ factor of the decomposition ${\bf U}_\sigma = Q D R$ contains information on the numerical scales in ${\bf U}_\sigma$, and can be used to truncate the model space by omitting the eigenspace of ${\bf U}_\sigma$ that corresponds to unoccupied states. At temperatures of interest this reduces the model space dimension to the order of a few hundreds and speeds up substantially the calculation of observables.

\subsection{Canonical ensemble observables}\label{results}

We have studied a number of observables for a finite-size trapped Fermi gas~\cite{Gilbreth2013} and for the uniform Fermi gas~\cite{Jensen2018} to better understand their pairing correlations and thermodynamics across the superfluid phase transition.

\subsubsection{Condensate fraction and critical temperature}
The condensate fraction describes the extent of off-diagonal long-range order (ODLRO) in the two-body density matrix~\cite{Yang1962} 
$\rho_2(i \uparrow ,j\downarrow; k \uparrow,l\downarrow)= \langle\hat{a}^\dagger_{i \uparrow}\hat{a}^\dagger_{j \downarrow}\hat{a}_{l \downarrow}\hat{a}_{k \uparrow}\rangle$.
For the trapped gas the indices $i,j,k,l$ refer to harmonic oscillator states, and for the uniform Fermi gas they refer to momentum states. When ODLRO is present, $\rho_2$ will acquire a maximal eigenvalue $\lambda_2$ which scales with the system size and corresponds to the occupation of a pair state. The value of $\lambda_2$ is bound by $[N(M-N/2+1)/(2M)] \leq N/2$, where $M$ is the number of single-particle states ($M = N_L^3$ for a lattice). The condensate fraction can then be defined as $n=\langle \lambda_{2} \rangle/[N(M-N/2+1)/(2M)]$ or $n=\lambda_{2}/(N/2)$. 

\begin{figure}[h]
\begin{minipage}[t]{0.45\textwidth}
\centering
\includegraphics[width=.9\linewidth]{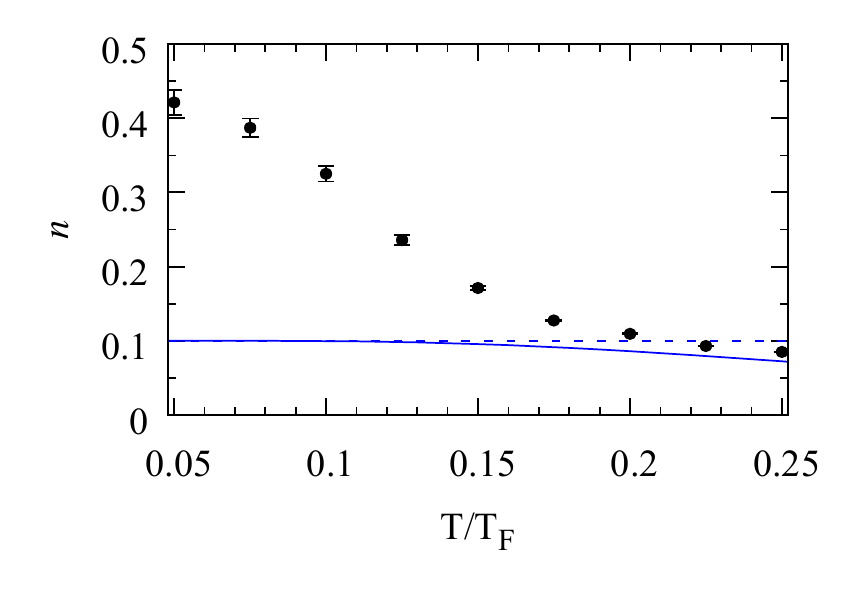}
\caption{Condensate fraction $n$ for a trapped system with $N=20$ particles as a function of $T/T_{F}$. The solid circles are the AFMC results, the solid line describes the noninteracting case, and the dashed line is the zero-temperature noninteracting limit $n=2/N$. Adapted from Ref.~\cite{Gilbreth2013}.}
\label{fig:ODLRO_trapped}
\end{minipage}
\hspace{0.25cm}
\begin{minipage}[t]{0.45\textwidth}
\centering
\includegraphics[width=1.0\linewidth]{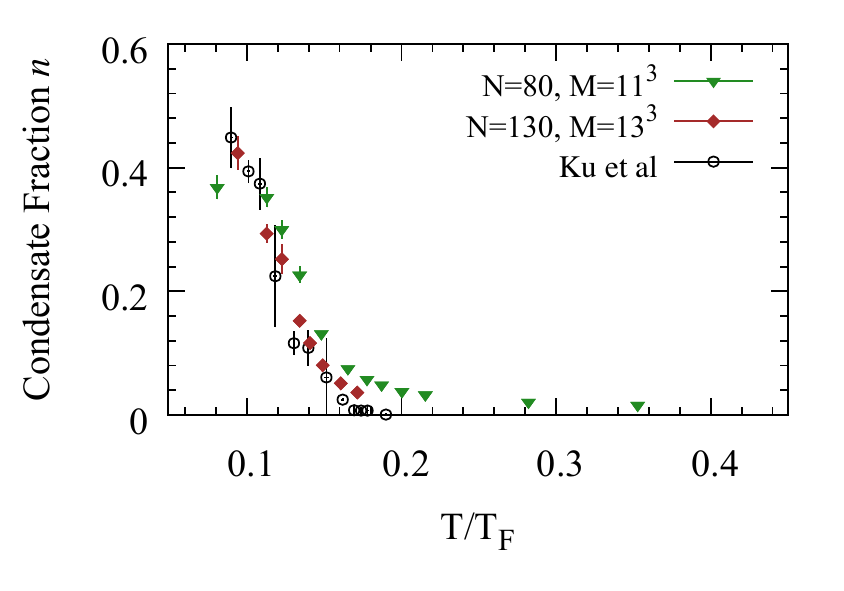}
\caption{Condensate fraction $n$ for the uniform gas vs.~$T/T_F$. AFMC results~\cite{Jensen2018} (solid symbols) are compared to the experimental result of Ref.~\cite{Ku2012} (open circles).  The agreement between the experimental result and the $N=130$ result is remarkable, but we note that our data is expected to vanish above our $T_c \sim 0.13 \,T_F$ in the limit of large lattices due to finite-range effects.}
\label{fig:ODLRO}
\end{minipage}%
\end{figure}

In Fig.~\ref{fig:ODLRO_trapped} we show  the condensate fraction $n=\lambda_{2}/(N/2)$ for a finite-size trapped gas with $N=20$ particles and  for a maximal number of oscillator shells $N_{\rm max}=11$, which is sufficient to reach convergence in $N_{\rm max}$ at $T=0.125 \, T_F$~\cite{GilbrethThesis}. For the trapped case $T_F = \epsilon_F = 4 \hbar \omega$ (using units with $k_B=1$). The condensate fraction begins to rise steeply below $T \approx 0.175 \, T_F$, showing a signature of the superfluid phase transition, which in a finite-size system is more precisely a smooth crossover. 

In Fig.~\ref{fig:ODLRO} we show the condensate fraction $n=\langle \lambda_{2} \rangle/[N(M-N/2+1)/(2M)]$ for the uniform gas with $N=80, 130$ particles on $11^3,13^3$ lattices, and compare to the experimental result of Ref.~\cite{Ku2012}.

For the uniform gas we use the condensate fraction to estimate the critical temperature, a procedure which requires finite-size scaling~\cite{Binder1981,Goldenfeld1992,Burovski2006,Goulko2010}.  In a preliminary analysis, we find  $T_{c}=0.130(15)~T_{F}$ at our finite density $\nu\simeq0.06$.  This density corresponds to a sizable ratio of the effective range to the Fermi wavelength $k_{F}r_{e}\simeq 0.41$, where $r_{e}=0.337\,\delta x$ for the lattice model simulated~\cite{Werner2012}. The zero-density limit results of Ref.~\cite{Burovski2006} and Ref.~\cite{Goulko2010} are given by $T_{c}=0.152(7)\,T_{F}$ and $T_{c}=0.173(6) \, T_{F}$, respectively. Extrapolating to zero density using AFMC would be useful in future work to remove the finite-range contribution. 

\subsubsection{Heat capacity}
\begin{figure}[h!]
\begin{minipage}[t]{0.45\textwidth}
\centering
\includegraphics[width=.9\linewidth]{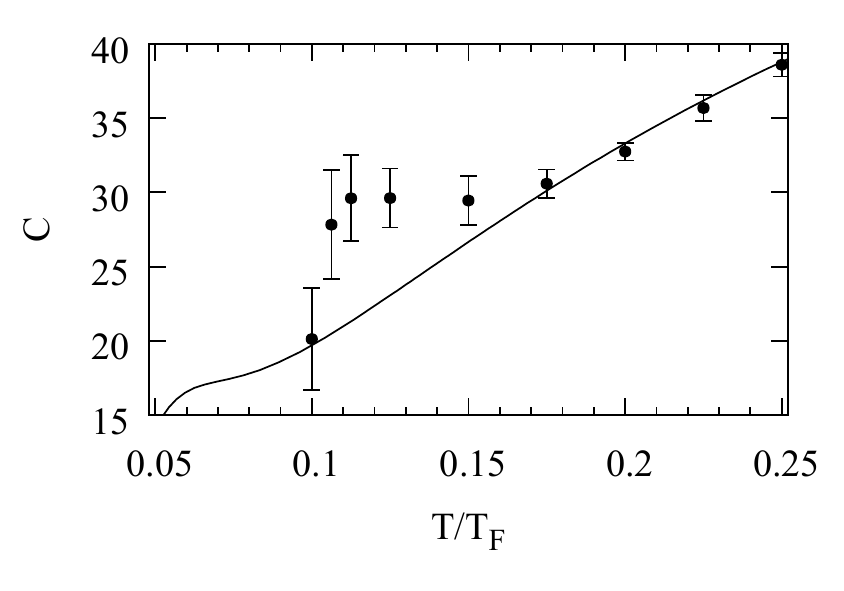}
\caption{Heat capacity for a trapped system with $N=20$ particles as a function of $T/T_{F}$. Symbols and and line are as in Fig.~\ref{fig:ODLRO_trapped}. Adapted from Ref.~\cite{Gilbreth2013}.}
\label{fig:HC_trapped}
\end{minipage}
\hspace{0.25cm}
\begin{minipage}[t]{0.45\textwidth}
\centering
\includegraphics[width=1.0\linewidth]{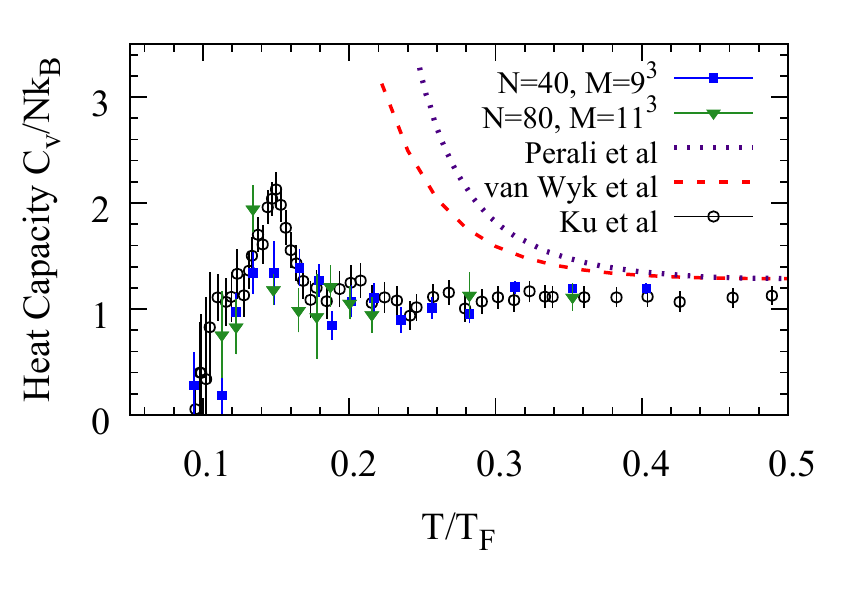}
\caption{AFMC heat capacity per particle~\cite{Jensen2018} (solid symbols) vs.~$T/T_{F}$ compared with the experiment of Ref.~\cite{Ku2012} (open circles) as well as the $T$-matrix result of Ref.~\cite{Strinati2011} (dotted line) and the NSR result of Ref.~\cite{Ohashi2016} (dashed line).}
\label{fig:HC}
\end{minipage}
\end{figure}

To compute the heat capacity we use the method of Ref.~\cite{Alhassid2001}, which substantially reduces the statistical errors by using the same auxiliary-field configurations to compute $E(T+\Delta T)$ and $E(T-\Delta T)$ when calculating $(\partial E/\partial T)_V$ numerically, taking into account correlations in the statistical fluctuations.
In Fig.~\ref{fig:HC_trapped} we show our result for the trapped case with $20$ particles and $N_{\rm max}=11$, for which $C$ is converged at $T \leq 0.2\,T_F$.
 This heat capacity exhibits a clear signature of the superfluid phase transition.
 
In Fig.~\ref{fig:HC} we show our result for the heat capacity of the uniform gas with $N=40$ and $N=80$ particles on lattices of size $M=9^3$ and $M=11^3$, respectively, and compare to experiment~\cite{Ku2012}, NSR theory~\cite{Ohashi2016}, and a non-self-consistent $T$-matrix calculation~\cite{Strinati2011}.  The results of Ref.~\cite{Ohashi2016} and Ref.~\cite{Strinati2011} seem to differ significantly from the experimental result and from our AFMC results. However, we note that the agreement between these results and experiment is better when plotted as a function of $T/T_{c}$, accounting for the differing values of $T_c$.  Our AFMC results agree well with experiment, allowing for a lower $T_c$ in the calculation, likely due to the finite density simulated.

\subsubsection{Pairing gap}
We have computed the energy-staggering pairing gap $\Delta_{E}$ in the canonical ensemble using Eq.~\eqref{gap} for both the trapped and uniform gases. $\Delta_E$ has the advantage of providing a model-independent signature of pairing without the need for analytic continuation of Monte Carlo results.

\begin{figure}[h]
\centering
\begin{minipage}[t]{0.45\textwidth}
\centering
\includegraphics[width=.9\linewidth]{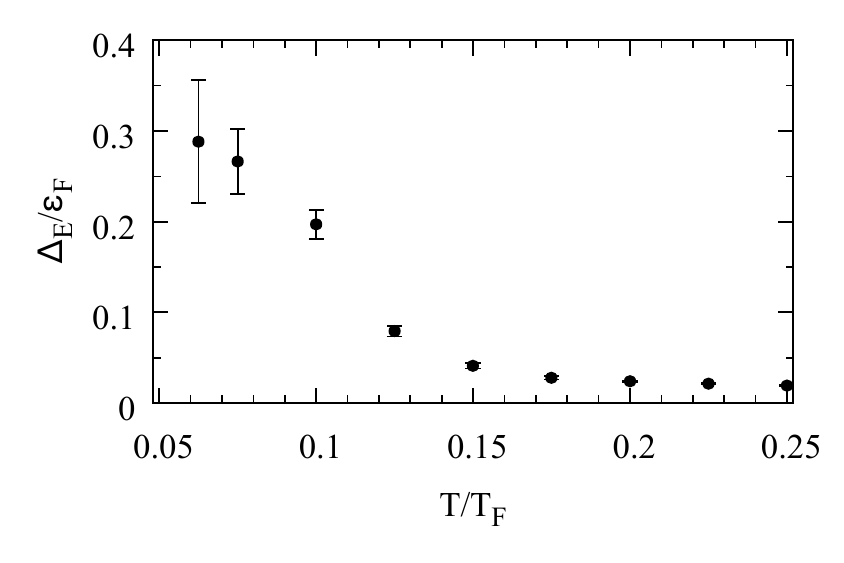}
\caption{Energy-staggering pairing gap $\Delta_E$ (in units of $\varepsilon_F$) for the trapped system with $20$ particles vs. $T/T_F$, computed using canonical-ensemble AFMC. Adapted from Ref.~\cite{Gilbreth2013}.}
\label{fig:gap_trapped}
\end{minipage}
\hspace{0.25cm}
\begin{minipage}[t]{0.45\textwidth}
\centering
\includegraphics[width=1.0\linewidth]{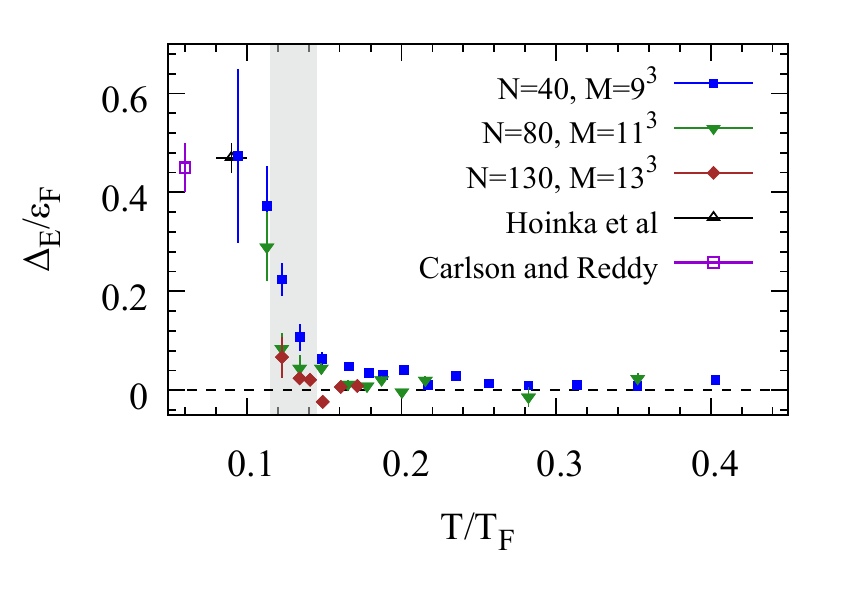}
\caption{Energy-staggering pairing gap $\Delta_E$ (in units of $\varepsilon_F$) for the uniform gas vs.~$T/T_F$ using canonical-ensemble AFMC. The shaded vertical band shows our preliminary estimate of $T_{c}$ calculated from the condensate fraction data using finite-size scaling.  Also shown are the $T=0$ quantum Monte Carlo result of Ref.~\cite{Carlson2008} (open square) and the recent experimental result of Ref.~\cite{Hoinka2017} (open triangle). Adapted from Ref.~\cite{Jensen2018}.}
\label{fig:Gap}
\end{minipage}
\end{figure}

In Fig.~\ref{fig:gap_trapped} we show $\Delta_E$ for the trapped case for $N=20$ and $N_{\rm max}=9$; it is converged in $N_{\max}$ for $T \leq 0.2 ~T_F$. The gap is slightly larger than zero at high temperatures due to finite-size effects. It begins to rapidly increase below $T \sim 0.175~T_F$, the same temperature at which features in $n$ and $C$ appear, showing no evidence of a gap prior to condensation for this finite number of particles.

In Fig.~\ref{fig:Gap} we show $\Delta_E$ for the uniform UFG for $N=40, 80, 130$ on lattices of size $M=9^3,11^3,13^3$, respectively.  We also show the $T=0$ quantum Monte Carlo result of Ref.~\cite{Carlson2008} and the low-temperature the experimental result of Ref.~\cite{Hoinka2017}.  These results are consistent with the spatially resolved radio-frequency spectroscopy result $\Delta_{E}/\varepsilon_{F}=0.44(3)$ of Ref.~\cite{Schirotzek2008}.  Our energy-staggering pairing gap $\Delta_{E}$ vanishes (or is very weak) for temperatures greater than the critical temperature $T_{c}\approx0.13~T_{F}$ of the finite density system ($\nu=0.06$).
\subsubsection{Spin susceptibility}
\begin{figure}[h]
\centering
\begin{minipage}[t]{0.48\textwidth}
\centering
\includegraphics[width=1.0\linewidth]{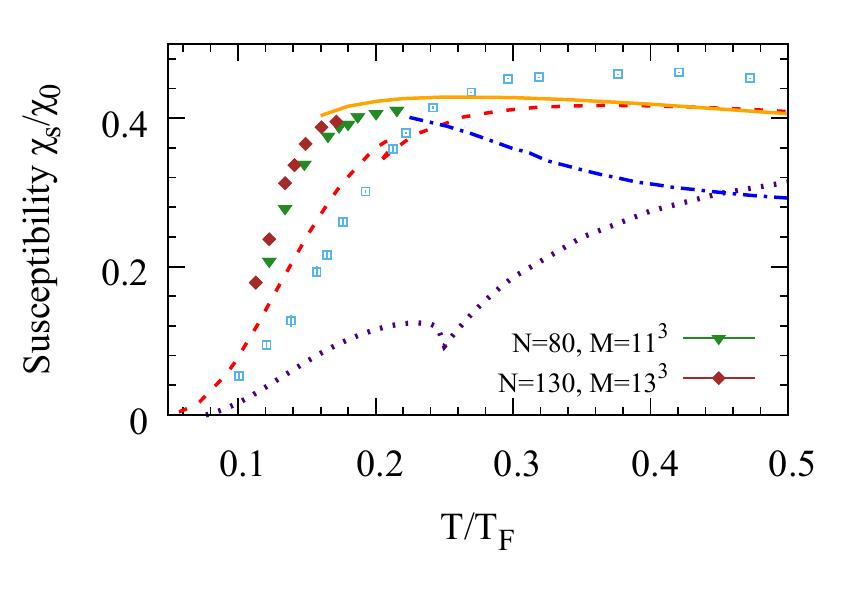}
\caption{Spin susceptibility $\chi_{s}$ for the uniform gas computed using our canonical-ensemble AFMC (solid symbols), the AFMC result of Ref.~\cite{Drut2013-2} (open squares), the $T$-matrix result of Ref.~\cite{Strinati2012-2} (dotted line), the $T$-matrix result of Ref.~\cite{Tajima2014} (dashed line), the fully self-consistent Luttinger-Ward result of Ref.~\cite{Enss2012} (solid line), and the self-consistent NSR result of Ref.~\cite{Pantel2014} (dashed-dotted line).}
\label{fig:chi}
\end{minipage}
\hspace{0.25cm}
\begin{minipage}[t]{0.48\textwidth}
\centering
\includegraphics[width=1.0\linewidth]{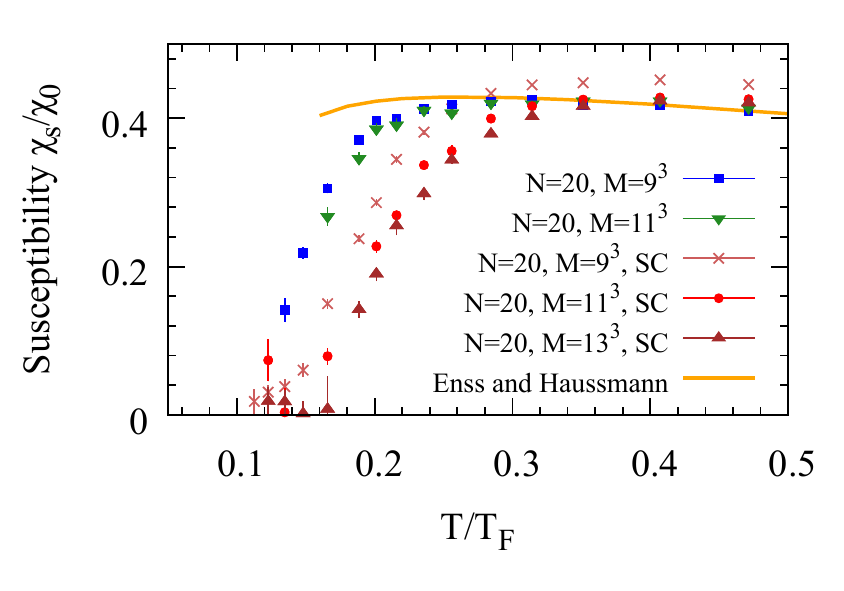}
\caption{AFMC results for the spin susceptibility of $\textrm{N}=20$ particles calculated with no cutoff on lattice sizes $9^3, 11^3$ and with a spherical cutoff (SC) on lattice sizes $9^3, 11^3, 13^3$.  There is a substantial difference between the no cutoff and spherical cutoff results as the continuum is approached for lower densities. The solid line is the result of Ref.~\cite{Enss2012}.} 
\label{fig:chi_compare}
\end{minipage}
\end{figure}

In Fig.~\ref{fig:chi} we show our AFMC results for the spin susceptibility $\chi_s$ of the uniform gas in units of the $T=0$ free Fermi gas susceptibility $\chi_{0}=3 \nu/2\varepsilon_F$, along with several other theoretical results. For this observable we use a single particle-number projection onto the total number $N = N_\uparrow + N_\downarrow$ of particles.  As discussed in Sec.~\ref{obs}, a downturn in the spin susceptibility is expected as the temperature is lowered below $T^{*}$.  Comparing the temperature scale where this downturn occurs to the critical temperature $T_{c}$ allows a determination of the pseudogap regime~\cite{Trivedi1995}.  The strong-coupling calculations of Refs.~\cite{Tajima2014,Strinati2012-2} are consistent with pseudogap physics, showing a "spin-gap" emerging via suppression in $\chi_{s}$ for $T>T_{c}$.  The AFMC result of Ref.~\cite{Drut2013-2} also shows clear suppression in $\chi_{s}$ for temperatures below $\sim 0.25\,T_F$, a value significantly higher than their estimated $T_{c}\lesssim 0.15(1) \, T_{F}$~\cite{Bulgac2008}.  The result obtained in the self-consistent Luttinger-Ward theory of Ref.~\cite{Enss2012} does not show a clear signature of a pseudogap and is consistent with the spectral function result for $T=T_{c}$ shown in Fig.~\ref{fig:Haussmann_spectral}.  The calculation of Ref.~\cite{Pantel2014}, which uses a modified and more self-consistent NSR approach, also shows no clear indication of a pseudogap or downturn in the spin susceptibility as the temperature is lowered. 

Our results show at most a weak signature of a spin-gap above $T_c$. We observe a downturn in the spin susceptibility for $N=130$ particles below $T\approx 0.17~T_{F}$, which is greater than our estimated $T_{c}$.  However, while we estimated $T_c$ in the thermodynamic limit for a density $\nu=0.06$ using finite-size scaling, no such thermodynamic limit has been taken with $\chi_{s}$, and the trend for larger systems seems to indicate that the effect will be smaller in this limit. 

\subsection{Spherical cutoff}\label{cutoff}

A spherical cutoff in the single-particle momentum of the uniform gas does not reproduce the two-particle scattering properties of the unitary gas even in the continuum limit~\cite{Werner2012,Jensen2018}.  The spin susceptibility provides a useful observable for testing the effects of a spherical cutoff on the many-particle physics.
In Fig.~\ref{fig:chi_compare} we show the spin susceptibility for a fixed number $N=20$ of particles on several lattice sizes, with and without a spherical cutoff in the single-particle momentum space.  Even at the lowest densities studied, $\nu=0.015, 0.009$ (corresponding, respectively, to lattice sizes of $11^3, 13^3$), the spherical cutoff effects survive and affect predictions for the value of $T^*$ and pseudogap physics.  

\section{Conclusion}\label{conclusion}
The nature of the unitary Fermi gas above the critical temperature $T_c$ for superfluidity has attracted much interest, both theoretically and experimentally. Some of the measured  thermodynamic observables are consistent with Fermi liquid behavior above $T_c$ 
(e.g., the pressure, Fig.~\ref{fig:expt_thermo}), while others are not (e.g., the heat capacity). Spectroscopy experiments have seen signatures of quasiparticles and pairing above $T_c$ (Fig.~\ref{fig:Jin2015_spectral}). Results of theoretical calculations still vary considerably, with some predicting  pronounced pseudogap effects and others not.
 In particular,  the heat capacity (Fig.~\ref{fig:HC}) and the  spin susceptibility (Fig.~\ref{fig:chi_compare}) vary widely among different theories.  
 
Measurements of the temperature dependence of the spin susceptibility and pairing gap would shed light on the pseudogap regime of the unitary gas.  Past experiments have used trapped systems to infer the properties of the uniform gas, but an experimental setup of a uniform gas was recently reported~\cite{Zwierlein2017}. Such experiments, expected in the near future, will address more directly the properties of the uniform gas and might provide better insight into pseudogap physics.  

Quantum Monte Carlo methods have provided accurate calculations of the critical temperature and thermal energy, but results for the pseudogap are either less reliable or incomplete. Within the framework of AFMC, a reliable extrapolation for both the thermodynamic and continuum limits would be a very significant achievement, particularly if the spectral function could be computed using the full first Brillouin zone of the lattice. This would provide a benchmark for strong-coupling theories and other methods, such as the bold diagrammatic Monte Carlo, whose convergence properties are not well known. Our AFMC calculations have made progress toward this goal by carefully fixing the density, studying convergence, computing new observables (the energy-staggering pairing gap and heat capacity), and resolving issues with the single-particle model space. However, AFMC calculations on larger lattices are still computationally very challenging.

\paragraph{Acknowledgments}  We thank N. Navon and G.F. Bertsch for useful discussions.  We also thank M.J.H. Ku for providing the experimental data of Ref.~\cite{Ku2012}, and  T. Enss,  P. Magierski, Y. Ohashi, P. Pieri, G. C.  Strinati, H. Tajima, M. Urban, and G. Wlaz\l{}owski for providing theoretical results shown in Figs.~\ref{fig:HC}, \ref{fig:chi}, and \ref{fig:chi_compare}. This work was supported in part by the U.S. DOE grant Nos.~DE-FG02-91ER40608 and DE-FG02-00ER41132.
The research presented here used resources of the National Energy Research Scientific Computing Center, a DOE Office of Science User Facility supported by the Office of Science of the U.S. Department of Energy under Contract No.~DE-AC02-05CH11231. It also used resources provided by the facilities of the Yale University Faculty of Arts and Sciences High Performance Computing Center.

\bibliographystyle{epj.bst}

\end{document}